# Distribution of the magnetic field relative to the plane of the Galaxy in the region of the Sagittarius spiral arm

**R.R. Andreasyan[1]; H.R. Andreasyan[1]; A. M. Mickaelian[1]; ; A.G. Suqiasyan[1] ;**

**G.M. Paronyan[1]**

1. Byurakan Astrophysical Observatory

randrasy@bao.sci.am

hasmik.andreasyan@gmail.com

andranik.suqiasyan.1995@mail.ru

paronyan_gurgen@yahoo.com **(corresponding author)**

*Abstract.* Faraday rotation data for 492 pulsars and more than 2,700 extragalactic radio sources were used to study the magnetic field, in detail, in the direction of Galactic longitude $33^0 < l < 63^0$, which partially covers the Sagittarius spiral arm region.

It was proposed that the Sagittarius spiral arm lies north of the h = -300 pc plane, where h is the distance from the Galactic plane. The (−) sign indicates that this plane is located in the southern hemisphere of the Galaxy. The magnetic field of the Sagittarius spiral arm is directed toward the Sun. This direction is close to the direction of the magnetic field of the northern hemisphere halo.

The small absolute value of the rotation measures in the region south of the h = -300 pc plane is due solely to the contribution of the magnetic field of the southern hemisphere halo, which is directed opposite to the field of the northern hemisphere halo and the direction of the field in the Sagittarius arm.

In the interarm region between the local Orion arm and the Sagittarius arm, the halo magnetic field extends to the Galactic plane, where a reversal of the magnetic fields of the halos of the northern and southern hemispheres occurs.

*1. Introduction:* The magnetic field of our Galaxy was discovered in 1949 [1] using observational data on the interstellar polarization of starlight. Studying the magnetic field of our galaxy, as well as other galaxies, is very important for understanding all the active processes occurring in galaxies. For example, the motion of cosmic rays in the Galaxy is entirely due to the presence of the Galaxy's magnetic field [2]. The magnetic field explains the synchrotron background radiation of the Galaxy [3]. It also has a regulating effect in the star-formation process [4]. Since the discovery of the Galactic magnetic field in 1949, it has been studied using observational data of various types, such as synchrotron radiation of the interstellar medium [5], Zeeman splitting of spectral lines of HI and different molecules in the radio range [6], polarized thermal emission of dust clouds, and Faraday rotation data for extragalactic radio sources and pulsars (for example [7-9]).

It should be noted that pulsars play a very important role in the study of the interstellar medium. Numerous baseline measurements are known for these objects, some of which are obtained directly from observations, others from the same observational data, but using different physical and morphological models of pulsars, as well as the medium through which polarized radio emission passes from the source to the observer. To study the Galaxy's magnetic field, data from the so-called Faraday rotation measure (RM) are used. The essence of the Faraday rotation effect is that when polarized radiation passes through a magneto-ionized medium, the plane of polarization of the radiation rotates. Moreover, the magnitude of the rotation depends on the wavelength of the radiation. This makes it possible to calculate the RM if polarization observations are available at several wavelengths using the following formula

$$RM = d\psi/d(\lambda^2). \qquad (1)$$

In this formula, $\psi$ is the rotation angle of the plane of polarization of the radiation, and $\lambda$ is the wavelength of the radiation.

On the other hand, RM theoretically depends on the electron concentration $n_e$ in $sm^{-3}$ and the projection of the magnetic induction $B_L$ in Gauss onto the line of sight in a magneto-ionized medium.

$$RM = \alpha\int n_e B_L dL \quad (\alpha=8.1 \cdot 10^5) \qquad (2)$$

Integration is performed along the radiation path from the pulsar to the observer in parsec.

Data on the dispersion measures (DM) of pulsars are also important for studying the interstellar medium of the Galaxy. The dispersion measure is determined from observations of pulsars using the effect that signal from the same pulse of pulsar at different frequencies $\omega$ reaches the observer at different times t. The difference in this time is related to the dispersion measure by the formula

$$t_2 - t_1 = (2\pi e^2/m)(1/\omega_2^2 - 1/\omega_1^2)DM , \qquad (3)$$

On the other hand, dispersion measure DM is related to the electron concentration $n_e$ in the interstellar medium by the following integral over the radiation path

$$DM = \int n_e dL . \qquad (4)$$

Using formula (4), one can find the distance between a pulsar and the Sun if the distribution of electron concentration $n_e$ in the Galaxy or at least its average value along the radiation path is known, also, using formula (2) together with formula (4), one can calculate the average value of the projection of magnetic induction in micro gasses (μG) onto the line of sight using the following formula

$$\langle B_L \rangle = (1/\alpha)(RM)/(DM) = 1.23(RM)/(DM). \qquad (5)$$

The first works devoted to the study of the magnetic field of the Galaxy using the data of pulsar rotation measures were carried out in the early 1970s, when the rotation measures of several tens of pulsars were known. At that time, $\langle B_L \rangle$ was estimated in the region of the local Orion arm [10]. In addition to an increase in the number of data of pulsar rotation measures, works appeared in which the magnetic field in more distant spiral arms and in regions at large distances from the plane of the Galaxy was studied. In particular, in works [8; 11], based on the analysis of data on the rotation measures of 185 pulsars and 802 extragalactic radio sources, a model of a two-component magnetic field of the Galaxy was proposed. According to this model, the flat component of the magnetic field of the spiral arms lies inside the component of the magnetic field of the Galactic halo, which has opposite directions in the northern and southern hemispheres of the Galaxy. Moreover, in the northern hemisphere, the direction of the halo magnetic field is opposite to the direction of the magnetic field of the local Orion arm.

In the paper by Han and Xu [12], they obtained the first quantitative estimate of the sizes of the huge magnetic toroids in the Galactic halo, which start at a Galactocentric radius of less than 2 kpc and extend to at least 15 kpc. The magnetic fields in the toroids in the northern and southern hemispheres of the Galaxy have opposite directions to each other. In this paper, we will study the dependence of the magnetic field in some selected directions on the distance from the plane of the Galaxy.

*2. Data used:* For the study, we use data from the ATNF pulsar catalogue 2025, as well as Faraday rotation data for 37,543 extragalactic radio sources from [13]. The ATNF pulsar catalogue 2025 contains various data for approximately 4,000 pulsars. In this catalogue, Faraday rotation measures are known for 1,990 pulsars. The data for these pulsars, as well as the Faraday rotation data for extragalactic radio sources, were used in this paper.

It should be noted that the rotation measures of pulsars are formed in three regions.

$$RM = RM_0 + RM_{ir} + RM_{in}$$

The first component $RM_0$ is the contribution from the circumpulsar region. The second component $RM_{ir}$ is formed in regions with irregular magnetic fields. The third component $RM_{in}$ is the contribution from the large-scale field of the Galaxy and is formed in the interstellar medium on the way from the pulsar to the observer.

The first component, which can be conventionally called the proper rotation measure of the pulsar, is considered negligible compared with the commonly observed RM values of pulsars. Finally, to estimate the proper rotation measure of pulsars, we can use data on the rotation measures of pulsars located in the same globular cluster. For these pulsars, the contribution from irregular and large-scale magnetic fields ($RM_{ir} + RM_{in}$) is the same and can be estimated as the average value of all rotation measures in this globular cluster. Thus, the proper rotation measure ($RM_0 = RM - RM_{ir} - RM_{in}$) for each pulsar located within a given globular cluster can be determined from observations. Data from globular clusters NGC104 (Tuc47), M15, NGC1851, and Terzan5, which contained at least 5 pulsars with known rotation measures, were used. It turned out that, in the first three clusters, the spread of RM values from their average value is negligible (the standard deviation $\sigma$ is no more than 2.5rad/m$^2$). For the globular cluster Terzan5, $\sigma$ is about 10 rad/m$^2$. Thus, when using pulsar rotation measures to study the large-scale magnetic field of the Galaxy, the contribution of the pulsar proper rotation measures can be neglected.

The turbulent random magnetic fields are much stronger than the large-scale magnetic field, but they have a coherence length of less than 100 pc [14]. Even if the electron density and magnetic-field strength in these regions are $n_e = 1$ cm$^{-3}$, B = 2μG, the contribution of irregular magnetic fields to the pulsar rotation measure ($RM_{ir}$) can be no larger than 200 rad/m$^2$. The contribution of irregular magnetic fields to the pulsar RM can be positive or negative, depending on the direction of the projection of this magnetic field onto the line of sight. In statistical analyses of RM data for pulsars, this value is considered a random component. It can be expected that the sign of RM at sufficiently large absolute RM values will not depend on the contribution of irregular magnetic fields. Therefore, to study large-scale magnetic fields in the selected directions, we will try to use data from pulsars with large absolute RM values (│RM│>200 rad/m$^2$, or │RM│>300 rad/m$^2$).

*3. The Galactic distribution of large absolute RM values of pulsars:* We used data from pulsars with large rotation measures to study the distribution of RM signs in the Galaxy magnetic fields. This makes it possible to identify clearly distinguishable regions of large-scale. It was constructed the distribution of rotation-measure signs of pulsars in the plane of the Galaxy. Were used 328 pulsars with |RM|> 300rad / m$^2$, and 496 pulsars with |RM|> 200rad / m$^2$.

It should be noted that in our earlier work [15], the distribution of fewer pulsars with rotation measures |RM|> 300rad / m$^2$, and |RM|> 200rad / m$^2$ in the Galactic plane was presented. It was shown that in the ring with Galactocentric distances of approximately 5 kpc <R<7 kpc, the magnetic field has a counterclockwise direction, while in the region R>7 kpc, the magnetic field has a clockwise direction. Note that the analysis of numerous new data on pulsar rotation measures fully confirms the above-mentioned conclusions.

In the work [16], it was shown that within the framework of the dynamo theory, the Galactic magnetic field can have so-called reversals associated with a direction change from one region to another. The reversals were modeled using the so-called no-z approximation, based on the fact that the Galactic disk is sufficiently thin. This was in good agreement with the result of work [15], obtained as a result of the analysis of observational data.

The distribution of RM signs in Figures 1 from the work [15] can also be interpreted as follows. The large-scale magnetic fields in two regions outlined by straight lines in this figure have different directions relative to the observer and may be associated with individual Galactic spiral arms. For a more detailed study of the above-mentioned regions, we plot the distribution of pulsars with large absolute values of rotation measures (|RM|> 200rad / m$^2$) as a function of the Galactic longitude (l) and the distance from the Galactic plane (h) (fig.1).

The analyses of data show that there are few pulsars with large RMs in the direction of the Galactic anticenter. Therefore, on the figure 1 we present the distribution of RM signs of pulsars toward the center of the Galaxy in coordinates $-90^0<l<90^0$. Note that, for clarity, the coordinates $l-360^0$ are given instead of Galactic longitudes $270^0<l<360^0$.

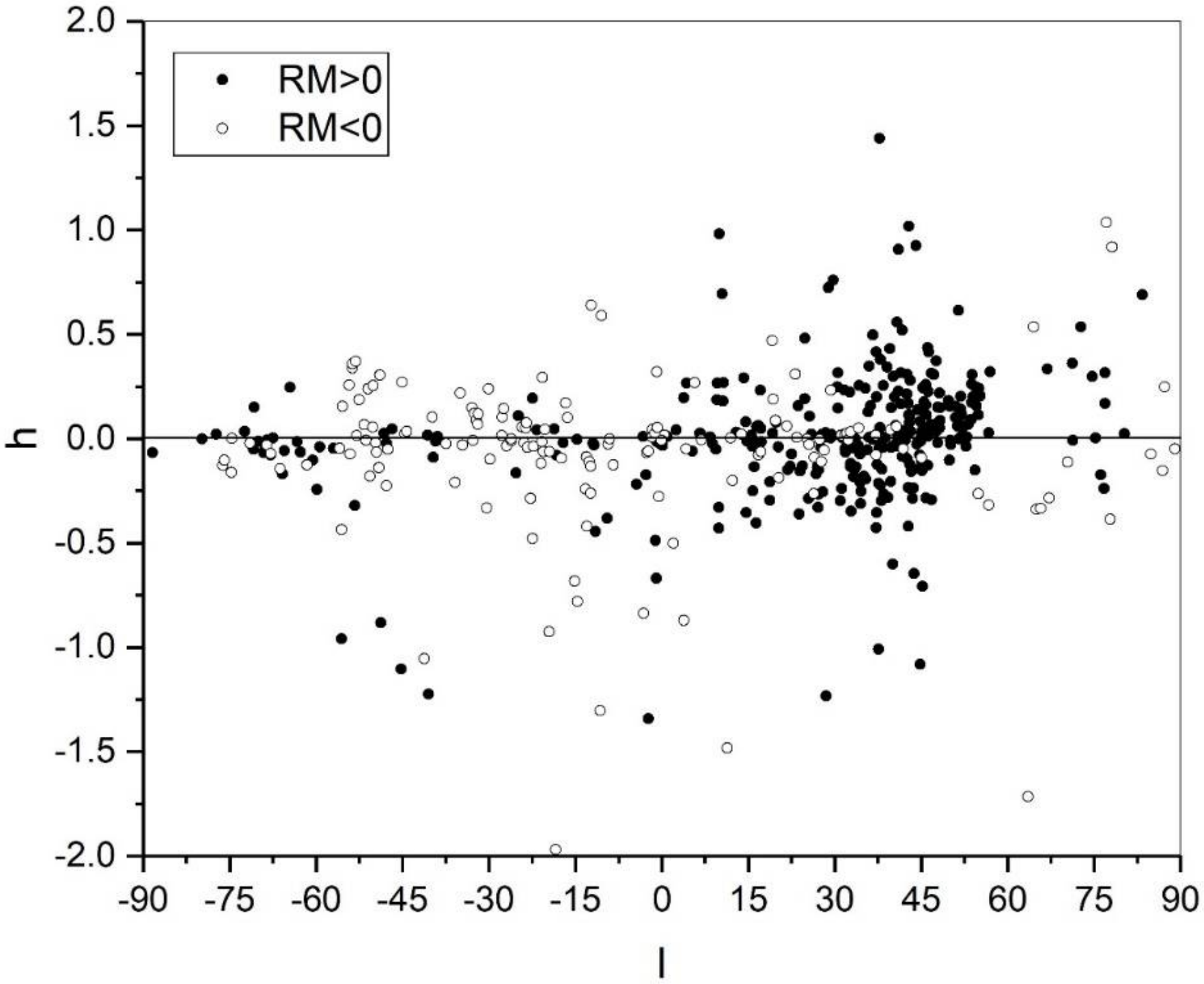

Fig.1. Distribution of Faraday rotation measures $|RM|>200$ rad/m$^2$ over galactic longitude l and the distance from Galactic plane h. Black circles denote pulsars in which RM> 200rad / m$^2$ (the magnetic field is directed to the observer), white circles denote pulsars in which RM < -200rad / m$^2$ (on the fig.1 for galactic longitudes $270^0<l<360^0$ we give the values of $l-360^0$).

In Figure 1 two areas are highlighted. In the first region, with Galactic longitudes of about $33^0 < l < 63^0$, where is the main part of the major Sagittarius spiral arm located (see the fig.11 of the work Georgelin & Georgelin [17]), almost all pulsars have positive rotation measures, whereas in the second region, $-56^0 < l < -26^0$, where is the main part of the intermediate Scutum-Crux spiral arm located, most pulsars (46 of 58) have negative rotation measures. Given that the Sagittarius spiral arm is located in the first region, we will study the magnetic field of this region in more detail.

In the figure 2 we present the distribution of pulsars with rotation measures |RM|> 200rad / m$^2$, located in the first region ($33^0 < l < 63^0$) in coordinates d and h, where d is the distance of the pulsar from the Sun.

Figure 2 shows that the region under study is dominated by pulsars with positive rotation measures (of the 173 pulsars with |RM|> 200 rad / m$^2$, only 9 have negative rotation measures). The pulsars are not symmetrically located relative to the Galactic plane. Most pulsars have h > -300 pc. Two southern hemisphere pulsars with large distances from the Galactic plane are located at distances from the Sun greater than 10 kiloparsecs and extend beyond the Sagittarius spiral arm.

There are no pulsars with |RM|> 200 rad/m$^2$ at distances closer than 2 kpc from the Sun.

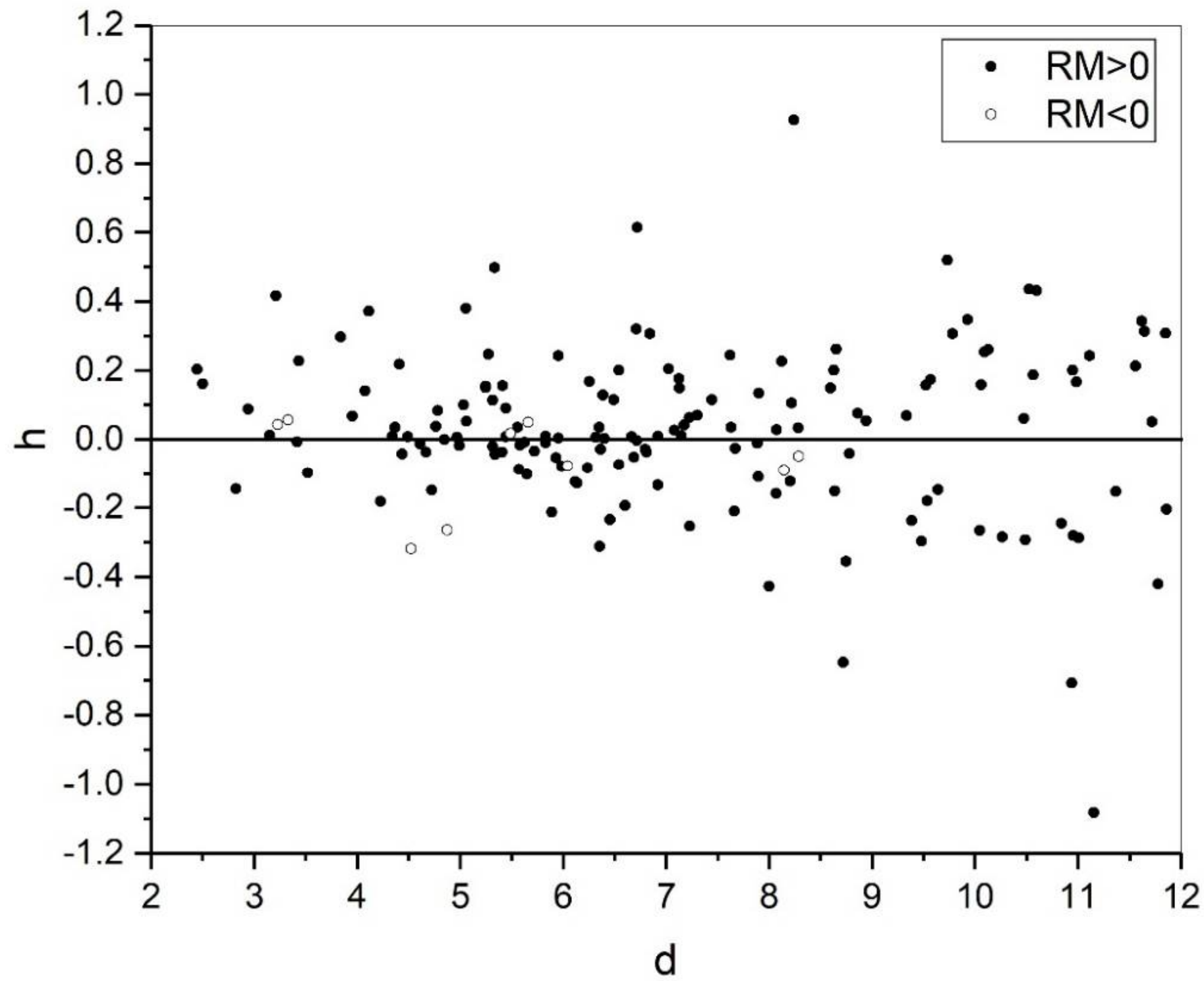

Fig. 2. The distribution of pulsars in coordinates d and h, where d is the distance of pulsar from the Sun and h is the distance from the Galactic plan. ($33^0 < l < 63^0$ and $|RM| > 200 rad/m^2$).

In the next paragraph, we will study the distribution of pulsars in the same region, but with $|RM| < 200 rad/m^2$.

*4. The Galactic distribution of pulsars with $|RM| < 200 rad/m^2$:* To study the distribution of pulsars with rotation measures $|RM| < 200 rad / m^2$ located in the region ($33^0 < l < 63^0$), we construct figure 3. The coordinates are d and h, as in the fig.2, but on this picture the mark $RM_{200}$ indicates that $100 rad/m^2 < |RM| < 200 rad/m^2$, and the mark $RM_{100}$ indicates that $0 < |RM| < 100 rad/m^2$.

Figure 3 shows that, as in figure 2, pulsars with positive rotation measures are primarily located above the line with h = -300 pc. In contrast to figure 2, there are lot of pulsars with negative rotation measures that are primarily located below the line with h = -300 pc. This means that the magnetic field direction south of this line is opposite to the field direction north of this line. Furthermore, at distances less than 2 kiloparsecs from the Sun, the transition line of the magnetic field direction coincides with the plane of the Galaxy. By analyzing the distributions of rotation measures shown in Figures 2 and 3, we can conclude that, at distances greater than 2 kiloparsecs from the Sun, the pulsar rotation measures are distributed as follows. Above the line with h = -300 pc, the rotation measures are mostly positive (the magnetic field is directed toward the Sun) and have large values on average (164 pulsars have RM> 200 rad / m$^2$), whereas below the line with h = -300 pc, the rotation measures are mostly negative (the magnetic field is directed away from the Sun) and have small absolute values on average (there are no pulsars with |RM|> 200 rad / m$^2$). We also recall that at distances from the Sun less than 2 kiloparsecs, the RM sign transition line practically coincides with the Galactic plane, and |RM|< 200 rad / m$^2$.

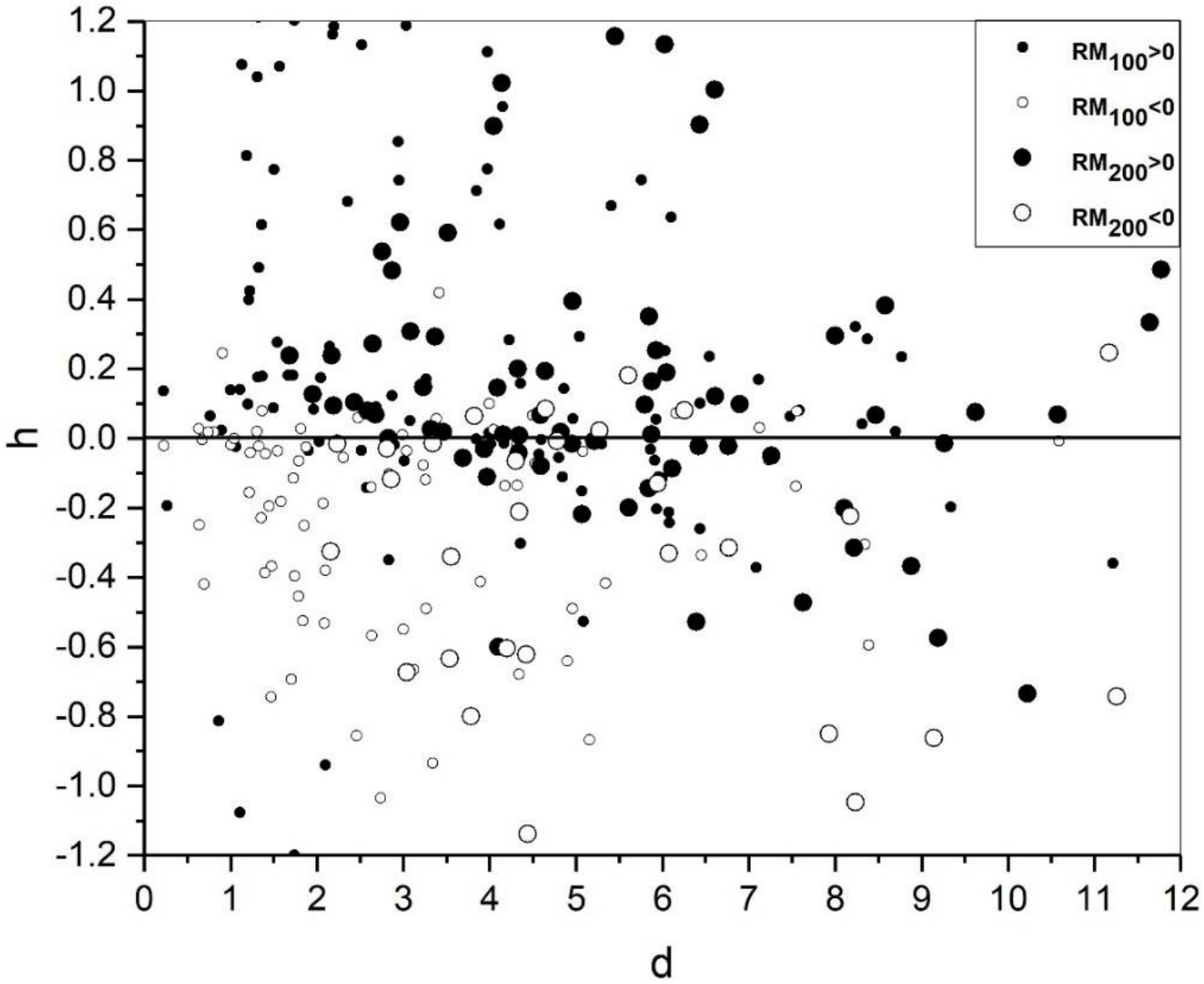

Fig. 3. The distribution of signs of pulsars with │RM│<200rad/m$^2$ from the region $33^0 < l < 63^0$.

The distribution of pulsar rotation measures in the region $33^0 < l < 63^2$, shown in Figures 2 and 3, fits well with the two-component magnetic field model of the Galaxy proposed earlier in [8; 11]. According to this model, the flat component of the magnetic field of the spiral arms is surrounded on all sides by the large-scale magnetic field of the Galactic halo. Moreover, the direction of the magnetic field in the northern-hemisphere halo is opposite to that in the southern-hemisphere halo. If this model is applied to the region studied in this paper and which contains most of the Sagittarius spiral arm, then the following can be concluded. The Sagittarius spiral arm is located above of the line h = -300 pc. The strong regular magnetic field in the Sagittarius spiral arm is directed toward the Sun. This direction is close to the direction of the magnetic field of the northern-hemisphere halo of the Galaxy. The combined contribution of these fields explains the large values of positive rotation measures above the line h = -300 pc. In the region south of the line h = -300 pc, only the oppositely directed magnetic field of the southern-hemisphere halo contributes to the pulsar rotation measures. This explains the small absolute rotation measures with negative signs.

The proposed model can also explain the distribution of rotation measures for nearby pulsars located at a distance of less than 2 kiloparsecs from the Sun (Fig. 3). It has been shown that the sign reversal of the rotation measures in this region occurs near the Galactic plane. The Sun is known to be located closer to the inner boundary of the local Orion Arm. The interarm distance between the Orion Arm and the Sagittarius Arm is approximately 2 kiloparsecs (see fig.4 in[17]). Thus, it appears that in the interarm region, the halo magnetic field extends to the Galactic plane, where the reversal of the magnetic fields of the northern and southern hemisphere halos occurs.

These explanations are also in good agreement with the model presented in the paper by Xu and Han (2024) [12], where they obtained the first quantitative estimate of the sizes of the huge magnetic toroids in the Galactic halo with reversed field directions above and below the Galactic plane.

To estimate the magnitude of magnetic induction in the Sagittarius spiral arm, graphs of the rotation measures versus dispersion measures (RM-DM) were constructed. It should be noted that plotting these graphs for different constraints on the size of the study region or rotation measures yields significantly different results. Figure 4 shows one of the RM-DM graphs for one of the acceptable constraints (RM>0; DM<450).
From formula 5, it follows that the average value of the magnetic induction component on line of sight can be calculated using the formula $\langle B_L \rangle = 1.23A$, where A is the linear approximation parameter of the RM-DM relationship. In Figure 4, this parameter is A=1.216, yielding $\langle B_L \rangle = 1.5\mu G$. For the various constraints mentioned, the value of parameter A varies within the range of 1.0-1.5, resulting in $\langle B_L \rangle$ being within the range of 1.23-1.845μG. The standard deviation σ of the magnetic induction was also estimated for $\langle B_L \rangle = 1.5\ \mu G$ (Fig. 4). This yields σ = 0.11.

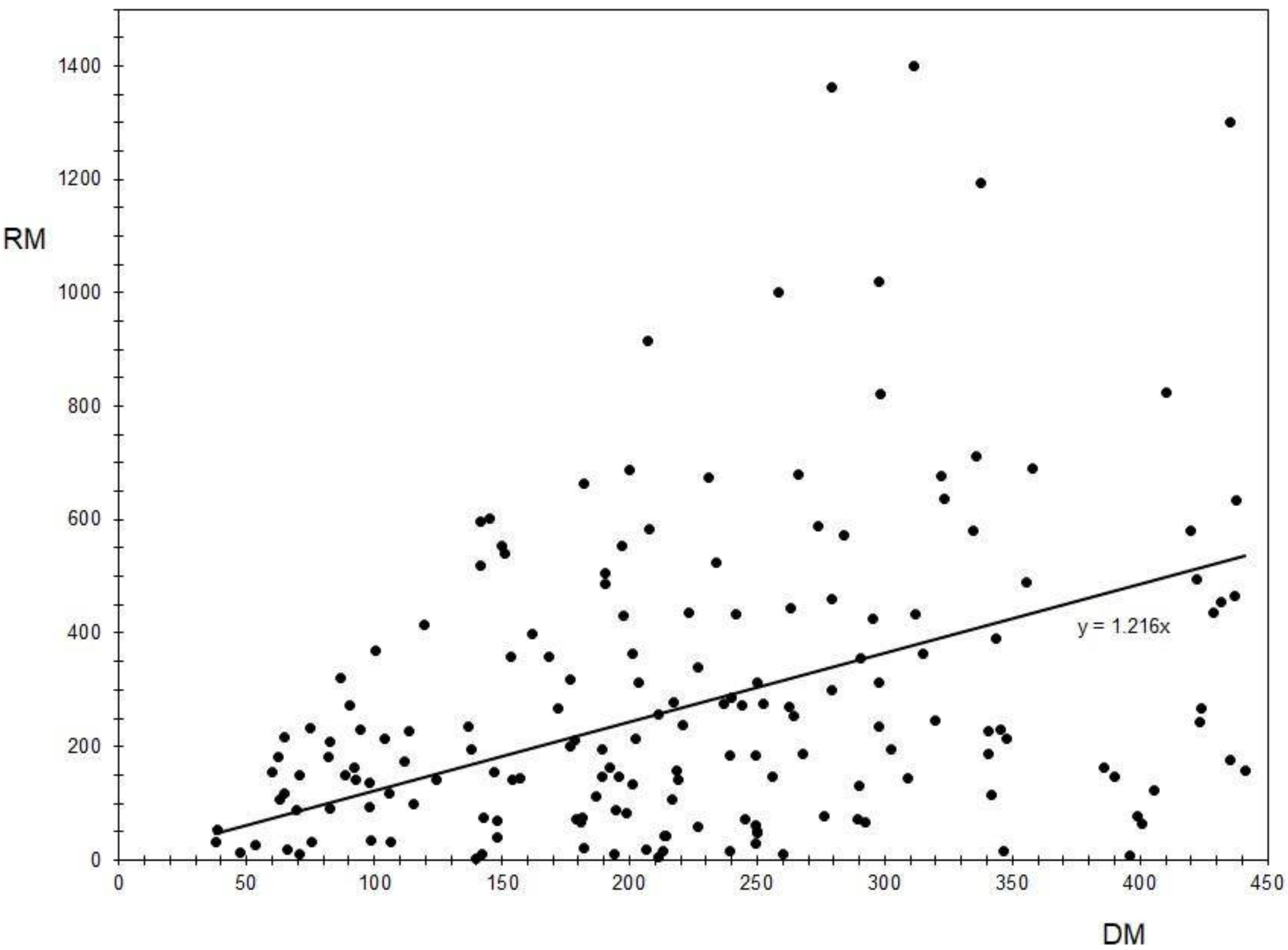

Figure 4. The relation of rotation measures from dispersion measures.

*5. The Galactic distribution of signs of RMs of extragalactic radio sources:* In [13], the NRAO VLA Sky Survey (NVSS) data were reanalyzed to derive rotation measures (RMs) toward 37,543 polarized radio sources. The resulting catalogue of RM values covers the sky area north of declination $-40^0$. We use the RM data from this catalogue to study the distribution of RM signs in the region $33^0 < l < 63^0$, which contains the Galactic Sagittarius spiral arm. In Fig.5 we present the distribution (in Galactic coordinates l; b) of rotation measures of more than 2700 extragalactic radio sources from [13] in the above-mentioned region of Galactic longitude.
To avoid overly cluttering the figure, pulsars with $|\mathrm{RM}| < 10$ rad/m$^2$ were excluded from the plot. This exclusion is partially justified by the fact that errors in calculating rotation measures sometimes reach such values.

Figure 5 clearly shows that in the aforementioned region, the signs of the rotation measures of extragalactic radio sources differ significantly between the northern and southern hemispheres of the Galaxy. While in the northern hemisphere RM is generally positive, in the southern hemisphere RM is negative.

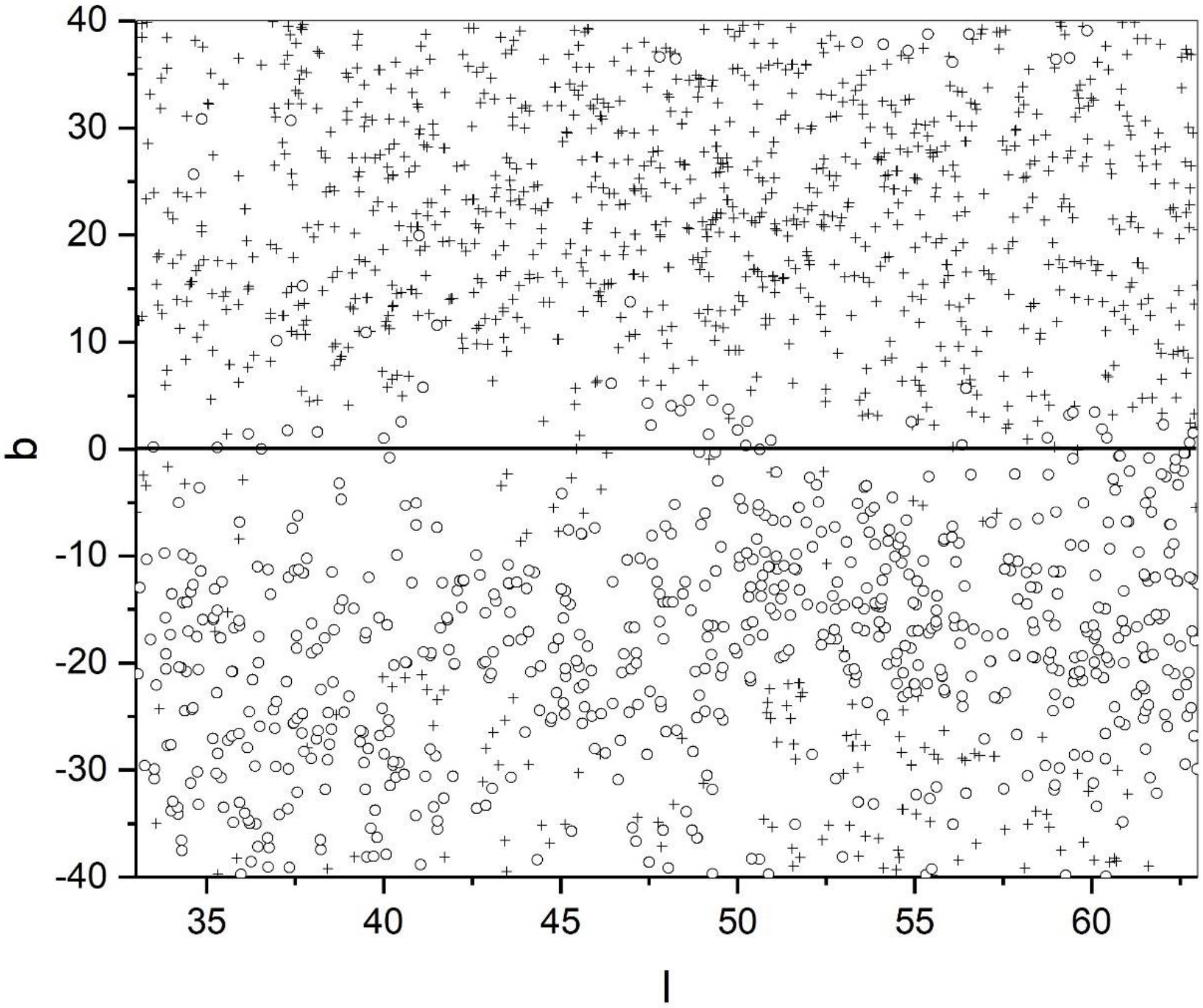

Fig.5 The distribution of rotation measure signs of extragalactic radio sources (in galactic coordinates (I; b)) in the galactic longitude region of $33^0 < l < 63^0$. The symbol + indicates, that RM is positive and white circle indicates that RM is negative.

In Table 1 we present the results of calculations of average rotation measures in different regions, divided by Galactic latitude in the Northern and Southern hemispheres of the Galaxy. For the calculations, we used rotation measures $|RM| > 10$ rad/m$^2$ of extragalactic radio sources from the region $33^0 < l < 63^0$.

Table 1

| b | $-5^0 \div 5^0$ | $5^0 \div 15^0$ | $15^0 \div 25^0$ | $25^0 \div 35^0$ | $35^0 \div 45^0$ | $45^0 \div 55^0$ |
|---|---|---|---|---|---|---|
| RM | -6.0 | +119.5 | +98.6 | +51.9 | +29.5 | +18.6 |

| b | $-5^0 \div 5^0$ | $-5^0 \div -15^0$ | $-15^0 \div -25^0$ | $-25^0 \div -35^0$ | $-35^0 \div -45^0$ | $-45^0 \div -55^0$ |
|---|---|---|---|---|---|---|
| RM | -6.0 | -94.0 | -57.0 | -10.7 | -0.7 | -0.1 |

From Table 1 we see that, at all latitudes in the Northern hemisphere of the Galaxy, the averaged rotation measures have positive signs, whereas in the Southern hemisphere they have negative signs. It also follows from Table 1 that at Galactic latitudes $b > 5^0$ the positive rotation measures in the Northern Hemisphere on average are much larger than the absolute values of the negative rotation measures of the Southern Hemisphere ($b < -5^0$). Such a distribution of signs of rotation measures is consistent with the aforementioned model of magnetic field in the above-mentioned region of galactic longitude. It was proposed that the regular magnetic field in the Sagittarius spiral arm is directed toward the Sun, and this direction is close to the direction of the magnetic field of the northern hemisphere halo of the Galaxy. Thus, the combined contribution of these two fields can explain the large values of positive rotation measures in the northern hemisphere. In the southern hemisphere, only the oppositely directed halo magnetic field contributes to the pulsar rotation measures. This explains the small absolute rotation measures with negative signs.

The very small average value of rotation measures in the region of Galactic latitudes $-5^0 < b < 5^0$ can be explained by the fact that the rotation measures of extragalactic radio sources are determined by magnetic fields along the entire path traversed by the radiation from the outer boundaries of the Galaxy to the Sun. At high Galactic latitudes, this radiation passes through the halo and, as a rule, through one spiral arm (in this case, the Sagittarius Arm). At low Galactic latitudes ($-5^0 < b < 5^0$), the radiation passes through all spiral arms along its path and, therefore, is determined by the contributions of the magnetic fields of all these arms, which can be opposite and compensate for each other. This explains the above-mentioned result outside the range of Galactic latitudes ($-5^0 < b < 5^0$).

*6. Conclusions.* Faraday rotation data on 492 pulsars and more than 2700 extragalactic radio sources were used in a detailed study of the magnetic field in the direction of galactic longitude $33^0 < l < 63^0$, which includes the Sagittarius spiral arm region.

The magnetic field in the region north of the plane h = -300 pc is well ordered and directed toward the sun. The magnetic field south of this plane has the opposite direction.

It was proposed that the Sagittarius spiral arm is located north of the plane h = -300 pc and that its magnetic field is directed toward the Sun and this direction is close to the direction of the magnetic field of the northern hemisphere halo. The combined contribution of these fields explains the large values of positive rotation measures above the line h = -300 pc.

The small absolute values of rotation measures in the region south of the plane h = -300 pc are caused only by the contribution of the halo magnetic field of the southern hemisphere of the Galaxy. This field is directed opposite to the field of the halo of the northern hemisphere.

In the interarm region between the local Orion arm and the Sagittarius arm, the halo magnetic field extends to the Galactic plane, where the reversal of the magnetic fields of the halos of the northern and southern hemispheres occurs.
It should be noted that the above-mentioned conclusions are confirmed by the analysis of data not only on the rotation measures of pulsars, but also on the distribution of rotation measures of extragalactic radio sources.

Thus, generalizing the results obtained in the present work for the Sagittarius spiral arm region and the previously obtained [8; 11] results for the Orion local arm region, it can be stated that both of these results fit well into the model of a two-component magnetic field of the Galaxy [11]. Moreover, the flat component of the magnetic fields of the spiral arms (in the present case, almost oppositely directed fields in the Orion and Sagittarius spiral arms) is surrounded by the magnetic field of the Galactic Halo, which has opposite directions in the northern and southern hemispheres of the Galaxy. Moreover, in the northern hemisphere, the direction of the Halo magnetic field is close to the direction of the magnetic field of the Sagittarius spiral arm and, therefore, is almost opposite to the direction of the field of the Orion arm. In the interarm region, the Halo field extends to the plane of the Galaxy, where a change in the direction of the field to the opposite occurs.

**Acknowledgments:** Advanced Research Grant 21AG-1C053: Revealing Early Stages of Galaxy Evolution Using Microwave Surveys of Active Galaxies, ANSEF 26AN:PS-astroth-3529